\documentclass[prc,aps,superscriptaddress,nofootinbib,floatfix,16pt]{revtex4-2}
\usepackage{graphicx} 
\usepackage[12pt]{extsizes}

\begin{document}

\title{\bf {\small {ELMO LIGHTS AND GLOW ON OBJECTS IN A CLOUD OF ELECTRICALLY CHARGED WATER DROPLETS}}}

\author{B.B.~Voitsekhovskii}
\affiliation{Hydrodynamics Institute, Novosibirsk, Russia \\
\vskip 0.125 in
published in Doklady Academy Nauk USSR, Fizika, 262 (1), 1982 \\ 
Communicated by academician A.N.~Skrinsky, May 18, 1981}
\maketitle
 
St.~Elmo's fire (SEF) is a rarely observed natural phenomenon that occurs in inclement weather when upward-pointing 
objects (sharp peaks of rocks, spires of towers, masts of ships) enter the lower negatively charged layer of clouds. 
Glowing brushes, torches—this is how SEF usually appears. 
The color is blue, sometimes red. 
The sizes of SEF can vary: from a few centimeters to one meter. 
Among the numerous observations of SEF (over a period of more than 20 centuries), the most mysterious cases 
are those of its appearance on raised hands, on people's heads, and on the crests of waves in the ocean.

The hypothesis of the electrical origin of SEF is present in the work of L.~Euler~\cite{Euler}. 
Until recently, SEF was observed exclusively in nature and identified with the well-known discharge from a sharp tip~\cite{BSE}, 
but many observations cannot be explained by a corona discharge, as the electric field strength at the flat surface of the Earth 
during a thunderstorm is only on the order of 100 V/cm~\cite{Chalmers}. 
In the work~\cite{PJETP}, for the first time under laboratory conditions, luminescence in a flow of charged droplets was detected. 
Here, the results of the study of the nature of the luminescence are presented.

Let's consider an object located in the electric field of a cloud of charged droplets. 
When droplets from the cloud settle on the object, a liquid film may form on its surface, which then separates into larger droplets. 
In this case, the conditions for the development of a discharge from the object are changing.

The phenomenon is based on the instability of the shape of liquid masses in an electric field discussed already by Gilbert~\cite{Gilbert}. 
The main results of the study of charged droplet instability were obtained by Rayleigh~\cite{Rayleigh}. 
The work~\cite{Taylor} shows that the shape of the surface of a conducting liquid can be stable if it forms a cone with a vertex angle of 98.6$^o$. 
With an increase in the electric field strength, a continuous jet of liquid emerges from the vertex of the cone~\cite{Zeleny}. 
In the case of viscous polarizable liquids, the jet may soon after emergence break into cylindrical pieces~\cite{Sedov}. 
Unlike natural conditions, where the electric field acts on droplets sitting on an object created by a cloud of charged droplets, 
the experiments~\cite{Taylor, Zeleny, Sedov} used the electric field of a system of conductors.

Discharges occurring under the influence of the electric field of a stream of charged droplets are also described in work~\cite{Barreto}, 
but the effects we are investigating were not observed there, possibly due to the high flow velocity at the observation site (more than 100 m/s).

\begin{figure}[ht!]
\begin{minipage}{0.47\linewidth}
\includegraphics[trim = 10mm 0mm 0mm 20mm, angle=0, width=0.98\linewidth]{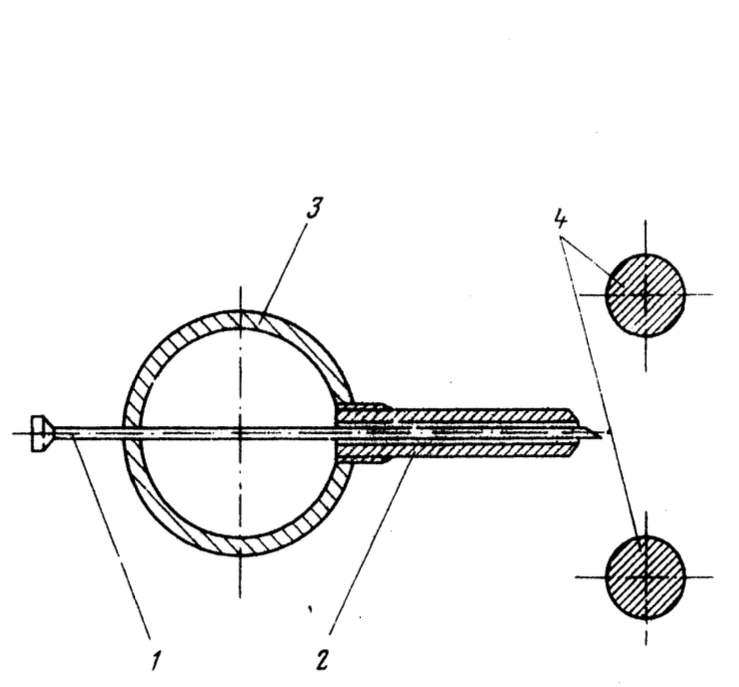}
\caption{Injector of the electrically charged droplets.}
\label{fig:Number1}
\end{minipage}
\hfill
\begin{minipage}{0.47\linewidth}
\includegraphics[trim = 0mm 0mm 0mm 0mm, angle=0, width=0.75\linewidth]{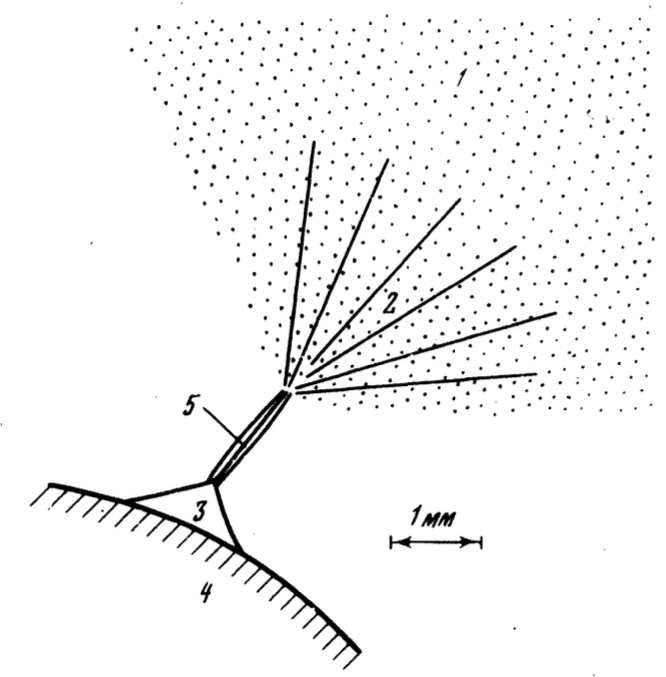}
\caption{Structure of fan-shaped luminescence. 1 - cloud, 2 - fan, 3 - droplet. 4 - grounded object, 5 - brightly glowing channel.}
\label{fig:Number2}
\end{minipage}
\hfill
\end{figure}

1. In our experiments, a generator of electrically charged droplets was used. It consists of 20 nozzles. 
Figure~\ref{fig:Number1} shows a cross-section of one of them. Medical needles (7) are used for liquid delivery. 
Air nozzles (2) are made of ebonite and are screwed into the wall of the tube (3), through which compressed air is supplied. 
The needles protrude from the nozzles by 2-3~mm. 
The nozzles are arranged in a single line along the tube at intervals of 12 mm. 
High voltage (10 - 20 kV) is applied to the electrodes (4). 
The needles and the tube (5) remain at zero potential relative to the ground.
The liquid emerging from the needle acquires a charge due to electrostatic induction. 
Streams of air flowing from the nozzles accelerate detachment of the droplets from the needles. 
Charged droplets, caught by the airflow, bypass the electrodes attracting them and form a cloud extending over 3 meters.
Near the generator, the cloud has a wedge shape. 
Its edge coincides with the line passing through the ends of the needles, and the angle of dispersion is 15-20 degrees. 
The experiments were conducted at a compressed air pressure of 2-7~atm and a liquid flow rate of $\sim$10 cm$^3$/s, air $\sim$20 nl/s.
The average droplet size in the cloud is 50~micrometers. 
Per second, the droplets carry about 100~micro~Coulombs into the cloud. 
Since the installation operates as an electro-hydrodynamic converter, isolated objects placed in the center of the cloud are charged to a high potential. 
Thus, the potential of a 1~cm diameter conductor reaches 70~kV relative to the Earth.

Let's note some features of the cloud generator here. It uses straight vertically positioned electrodes instead of the ring ones used in work~\cite{PJETP}. 
This allowed for an increased density of nozzle placement and eliminated the occurrence of breakdowns between the needles (7) and electrodes (4), which arise due to water accumulation on the horizontal sections of the ring electrodes. 
The electrodes are made from 10~mm diameter rods and are wrapped in black cotton thread. 
Before the experiment, the wrapping was soaked in water, which ensured that the surface of the electrodes was well-moistened and no droplets formed on it.

2. The main results were obtained through visual observations and photographing the glow on various grounded objects placed in a cloud of charged droplets. 
Two forms of glow were discovered. 
The first, known from the work~\cite{PJETP}, hereinafter referred to as the fan form, occurs on water droplets. 
Figure~\ref{fig:Number2} shows the structure of the fan in the case of a negatively charged cloud. 
At the base of the fan is a droplet, roughly cone-shaped.
From its tip, a brightly glowing channel begins, ranging from 3 to 5~mm in length and less than 0.5~mm in diameter. 
It transitions into a glowing fan.
When an object is placed in the center of a negatively charged cloud, the length of the fan reaches 5-8~cm, and the volume is 30-40 cm$^3$. 
When the object is located near the edge of the cloud but outside of it, the fan takes the shape of a cone with an angle of approximately 70$^o$ and a length of ~3~cm.
Depending on the position of the glow relative to the cloud and other grounded objects, the current flowing through the discharge varies from 1 to 30~$\mu$A. 
As the object is moved away from the cloud, the glow weakens, and when the current drops below 0.2-0.5~$\mu$A, the fan glow is no longer observed.
In a positively charged cloud, the fan glow is much weaker—the dimensions of the glowing area are 10 to 20 times smaller than in a negatively charged cloud.

The role of a water droplet located on the surface of an object in the formation of fan-shaped luminescence is clearly demonstrated in the following experiments. Firstly, it was found that fan-shaped luminescence is absent on hygroscopic bodies. It is also not observed at the end of a rod if its diameter is less than 0.5~mm, as no droplet forms at the end of the rod due to the blowing away of the liquid, which is associated with the luminescence. 
Secondly, it has been shown that on the same object, fan-shaped luminescence occurs when it is at a low temperature (around 0~$^o$C) and disappears when it is heated significantly.\\

\begin{figure}[ht!]
\begin{minipage}{0.58\linewidth}
\includegraphics[clip, trim = 14mm 50mm 0mm 20mm, angle=0, width=0.9\linewidth]{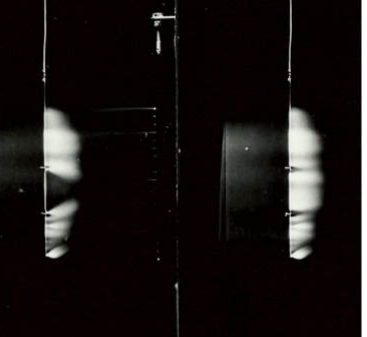}
\caption{Glow on heated (left) and cold (right) conductors. Air streams with droplets blow from right to left.}
\label{fig:Number3}
\end{minipage}
\hfill
\begin{minipage}{0.38\linewidth}
\includegraphics[clip, trim = 0mm 9mm 0mm 30mm, angle=0, width=1.\linewidth]{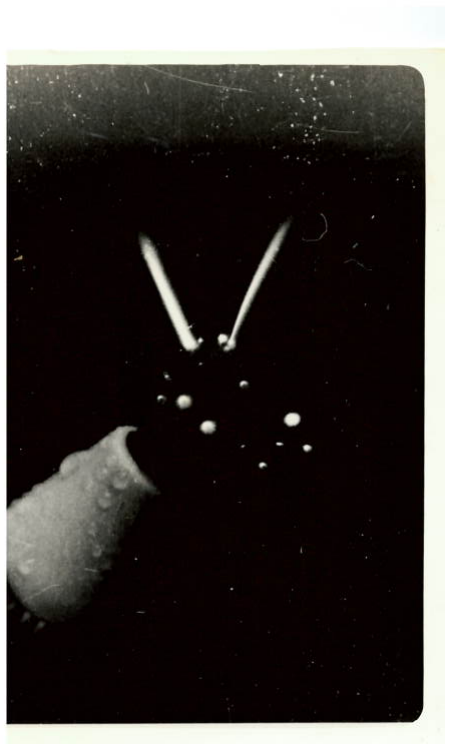}
\caption{Glow on a shaving brush.}
\label{fig:Number4}
\end{minipage}
\hfill
\end{figure}

Figure~\ref{fig:Number3} shows photographs of the glow on a conductor made of nichrome (in the center) and copper wire segments with a diameter of 0.5~mm.
When a current of ~10~A passes through the conductor, the central section (which, like the entire conductor, is in an airflow with droplets) heats up to 200-300~$^o$C. 
There is no glow on the hot section of the conductor, and there is no break in the glow on the cold conductor (see Fig.~\ref{fig:Number3}). 
The glow on the hot conductor is absent because the water that comes into contact with it from the cloud immediately evaporates. 
The fans from the cold sections of the conductor (see Fig.~\ref{fig:Number3}-left) are directed towards the central part of the cloud, 
whereas in Fig.~\ref{fig:Number3} they are directed along its axis: the fans discharge the cloud significantly faster than the hot wire.
In all experiments, the shielding effect is observed, which consists of the weakening of the glow when any grounded objects are placed between it and the charged droplet generator. 
The shielding is most clearly expressed with a negatively charged cloud, when a fan-shaped glow appears on the shielding object.\\

A significant interest is presented by the comparison of the fan-shaped glow and the observations~\cite{Macky}, in which behavior of freely falling water droplets in a region with a strong electric field was studied. 
When droplets with a diameter of 1-3~mm entered horizontal or vertical fields with an intensity of about 10~kV/cm, a glowing cone appeared. 
It started at the positively induced end of the droplet and extended to the negatively charged plate of the setup. 
In photographs taken with additional spark illumination, long jets emerging from the droplet can be seen. 
Apparently, the same phenomenon is observed in both the experiments~\cite{Macky} and in our case: breakdown along the surface of 
thin water streams emerging from the droplet.\\

In the latest series of experiments, luminescence was observed on a fine brush (Fig.~\ref{fig:Number4}). 
A remarkable feature of it is the cylindrical shape of the luminescent area, the stability of the orientation, and the shape of the luminescence. 
When the brush is inserted deeper into the cloud, the number of such "needles" increases.\\

The second form of luminescence—point like (p.l.)—occurs at the tips (see Fig.~\ref{fig:Number4}). 
Its characteristic feature is a small volume of luminescence, no more than 30~mm$^3$. 
When the tip is located at the center of the cloud, the luminescent area has a spherical shape, and a current of up to 40~$\mu$A flows through it. 
Such luminescence can occur at a greater distance from the cloud than the fan-shaped one. 
It does not change with the polarity of the cloud and disappears when the current through the tip is less than 0.05~$\mu$A. 
If an object with numerous tips (for example, wool fabric) is placed in the center of the cloud, p.l. can be observed on many 
fibers that extend along the electric field. 
On objects that are hygroscopic and do not have tips (such as the electrodes we made for the generator), luminescence is not detected, which may be related to its low brightness, as in this case, charged droplets from the cloud settle evenly over the surface of the object.\\

In conclusion, it should be noted that ball lightning and St.~Elmo's fire occur under similar conditions, and with an increase in the size of the cloud, it may be possible to experimentally investigate the nature of ball lightning. \\
Thus, we have established that: \\
1) a necessary condition for fan-shaped luminescence on an object is the presence of water droplets on its surface, which deform under the influence of the electric field of the volumetric charge cloud; any action that destroys the droplets (heating, absorption) leads to the disappearance of the luminescence; \\
2) on pointed objects (needles, bristles) placed in a cloud of charged droplets, fan-shaped luminescence does not occur, only small areas near the tips glow; \\
3) SEF and the fan-shaped luminescence are identical both in the conditions of their occurrence and in the shape and color of the luminescence.


\begin{thebibliography}{99}
% 
\bibitem{Euler} Euler~L.,~Letters~to~a~German~Princess~on~Various~Subjects~of~Physics~and~Philosophy,~St.-Petersburg,~1768. 
\bibitem{BSE} Bol'shaia Sovetskaia Encyclopedia, {\bf 30}, 503, 1974. 
\bibitem{Chalmers} Chalmers~J.A., Atmospheric Electricity. Pergamon Press, Oxford, 1967, \\
https://rmets.onlinelibrary.wiley.com/doi/10.1002/qj.49709540321.
\bibitem{PJETP} Voitsekhovsky B.V., Voitsekhovsky B.B., Pis'ma Zh.Eksp.Teor.Fiz, {\bf 23(1)}, 37 (1976), https://arxiv.org/pdf/2503.07652.
\bibitem{Gilbert} Gilbert W., On the Magnet and Magnetic Bodies, and on That Great Magnet the Earth, Book 2, 1600.
\bibitem{Rayleigh} Lord~Rayleigh,~Phil.~Mag.~{\bf~14},~184,~(1882),~https://doi.org/10.1080/14786448208628425.
\bibitem{Taylor} Taylor~G.,~Proc.~Roy.~Soc,~A~{\bf~280},~363~(1964),\\ https://royalsocietypublishing.org/doi/10.1098/rspa.1964.0151. 
\bibitem{Zeleny} Zeleny~J.,~Phys.~Rev.,~{\bf~10},~1~(1917),~https://journals.aps.org/pr/abstract/10.1103/PhysRev.10.1.
\bibitem{Sedov} Sedov~L.I.,~The~Thoughts~on~Science~and~on~Scientists,~Nauka,~Russian~Academy~of~Sciences,~Moscow,~1980.
\bibitem{Barreto} Barreto~E.,~Journal~Geophys.~Res.,~{\bf~74},~6911,~(1969),\\https://agupubs.onlinelibrary.wiley.com/doi/abs/10.1029/JC074i028p06911.
\bibitem{Macky} Macky~A.,~Proc.~Roy.~Soc,~A {\bf~133},~565~(1931),\\ https://royalsocietypublishing.org/doi/10.1098/rspa.1931.0168.
%
\end{thebibliography}
\end{document}